%% file: main.tex
\documentclass[12pt]{article}

\usepackage{sbc-template}
\usepackage{graphicx,url}
\usepackage[utf8]{inputenc}
\usepackage[brazil]{babel}
\usepackage{tabularx}

\title{IDEIA: A Generative AI-Based System for Real-Time Editorial Ideation in Digital Journalism}

\author{X\inst{1}}
\address{X}

\author{Victor B. Santos\inst{1}, Cauã O. Jordão\inst{1}, Leonardo J. O. Ibiapina\inst{1}, Gabriel M. Silva\inst{1},
\\Mirella E. B. Santana\inst{1}, Matheus A. Garrido\inst{1}, Lucas R. C. Farias\inst{1,2}}
  
\address{
  Análise e Desenvolvimento de Sistemas\\
  CESAR School - Recife, Pernambuco -- Brazil
  \email{\{vbs3, cof, ljoi, gms2, mebs, mag2, lrcf\}@cesar.school}
  \nextinstitute
  Departamento de Ciência da Computação\\
  Universidade Católica de Pernambuco (UNICAP) -- Recife, PE -- Brazil
  \email{lucas.farias@unicap.br}
}

\begin{document} 

\maketitle

\begin{abstract}
This paper presents IDEIA (Intelligent Engine for Editorial Ideation and Assistance), a generative AI-powered system designed to optimize the journalistic ideation process by combining real-time trend analysis with automated content suggestion. Developed in collaboration with the Sistema Jornal do Commercio de Comunicação (SJCC), the largest media conglomerate in Brazil’s North and Northeast regions, IDEIA integrates the Google Trends API for data-driven topic monitoring and the Google Gemini API for the generation of context-aware headlines and summaries. The system adopts a modular architecture based on Node.js, React, and PostgreSQL, supported by Docker containerization and a CI/CD pipeline using GitHub Actions and Vercel. Empirical results demonstrate a significant reduction in the time and cognitive effort required for editorial planning, with reported gains of up to 70\% in the content ideation stage. This work contributes to the field of computational journalism by showcasing how intelligent automation can enhance productivity while maintaining editorial quality. It also discusses the technical and ethical implications of incorporating generative models into newsroom workflows, highlighting scalability and future applicability across sectors beyond journalism.
\end{abstract}

Key-words: Digital journalism; Google Trends; Artificial Intelligence; Agile methodologies; Scrum; Editorial efficiency.

\input{text/body}

\bibliographystyle{sbc}
\bibliography{sbc-template}

\end{document}

%% file: text/body.tex
\section{Introduction}
\input{text/1_Introduction}

\section{Background}
\input{text/2_Background}

\section{Methods}
\input{text/3_Methods}

\section{Results}
\input{text/4_Results}

\section{Conclusion}
\input{text/5_Conclusion}

%% file: text/1_Introduction.tex
Contemporary technological advances have caused significant transformations in production structures, requiring continuous adaptations by professionals in multiple sectors. In communication, journalism has been particularly impacted by digitalization, the rise of social networks, and real-time information consumption. In this scenario, journalistic activity has come to demand speed in production and the ability to generate thematically relevant content that is visually attractive and sensitive to the fragmented and competitive digital context \cite{saenz2002ciencia}.

In light of this scenario, tools based on Artificial Intelligence (AI) emerge as strategic allies in the reinvention of editorial routines. In particular, generative AI offers solutions beyond data processing, actively contributing to creating textual, visual, or multimodal content and challenging limits previously attributed to exclusively human creativity \cite{amankwah2024impending}.

This paper presents the IDEIA project (Portuguese acronym for Inteligência Dedicada para Escrita e Inspiração Avançada), a solution developed in collaboration with the  Sistema Jornal do Commercio de Comunicação (SJCC), the largest media conglomerate in the North and Northeast of Brazil. The tool aims to optimize the journalistic ideation process by integrating real-time trend data through the Google Trends API with generative models based on the Gemini API \cite{el2024unleashing}. With this combination, the system can suggest topics, generate contextually relevant titles, and provide initial summaries for the writer, acting as assistive creative support \cite{choi2012predicting}.

The proposal of the IDEIA addresses relevant technical challenges, such as the orchestration of multiple APIs, ensuring interface responsiveness, and efficient manipulation of data in real-time. To this end, a modern architecture was adopted, with a back-end in Node.js, a front-end in React, and automated Continuous Integration (CI/CD) practices \cite{nadeem2022case}. This study aims not only to demonstrate the technical feasibility of the application but also to reflect on the role of computational solutions in improving digital editorial production, promoting a balance between human creativity and intelligent automation.

This paper is organized as follows. Section 2 presents the technical background of the project, describing the collaboration tools and technologies adopted in the development of the IDEIA system. Section 3 details the methodology, focusing on the application architecture, the development pipeline, and the technical challenges faced during implementation. Section 4 then discusses the results obtained, evaluating the practical impacts of the solution in the journalistic environment and highlighting the gains in editorial efficiency. Finally, Section 5 gathers the conclusions and proposes future directions, highlighting the contributions of the study to the integration of generative artificial intelligence in journalistic workflows.

%% file: text/2_Background.tex
The development of the IDEIA system required the careful selection of collaboration tools and technologies that would ensure efficiency, scalability, and cohesion between the different stages of the software life cycle. This section presents the primary resources adopted, grouped according to their function in the application construction process.

\subsection{Collaboration Tools and Development Support}
	
Collaborative management was conducted through \textbf{Trello}, using the agile methodology with Kanban boards structured in four columns (Backlog, To do, In Progress, and Done), allowing traceability and prioritization of tasks per sprint. For version control and continuous integration,\textbf{ Git and GitHub }were used, enabling distributed versioning and coordination of multiple developers with a complete history of commits and branches. Coding was primarily done in the IDE \textbf{Visual Studio Code}, chosen for its lightness, adaptability to multiple languages, and a broad ecosystem of extensions. The interface prototyping and alignment between the technical team and stakeholders were conducted with \textbf{Figma}, promoting early visual validation of flows, components, and information hierarchy.

\subsection{Front-end Technologies}
	
The presentation layer of the system was developed with \textbf{ReactJS}, a widely consolidated library for single-page application (SPA) reactive web applications, favoring modularity and component reuse. Using on-demand loading, the \textbf{Vite} packager was adopted to optimize build time and provide a more efficient development environment. \textbf{Styled-components} were used for styling, allowing CSS-in-JS encapsulation and cohesive maintenance of the visual style. Integration with REST APIs was implemented via \textbf{Axios}, ensuring robust asynchronous communication between the front and back-end. Finally, the Framer \textbf{Motion} library was used to add animated transitions and visual interactions, expanding the usability and aesthetic appeal of the application.

\subsection{Back-end Technologies}

The logical infrastructure of the application was built with \textbf{Node.js} and the \textbf{Express.js} framework, enabling a scalable and modular API. The JavaScript language was used homogeneously throughout the stack, facilitating the integration of the layers. As a differentiator, two external APIs were incorporated: the \textbf{Google Trends API}, which is responsible for providing updated data on trending terms, and the \textbf{Google Gemini API}, a generative model used to suggest titles and summaries. The persistent storage was structured in \textbf{PostgreSQL}, managed via a Docker container, and accessed through the \textbf{Sequelize ORM}, simplifying the manipulation of relational data in JavaScript. The application was containerized with \textbf{Docker}, allowing for environment isolation, replicability, and portability. To ensure transparency and interoperability in the development of the API, \textbf{Swagger} was used, a tool that generates automatic documentation and interactive interfaces for testing.

\begin{figure}[htbp]
\centering
\includegraphics[width=1\textwidth]{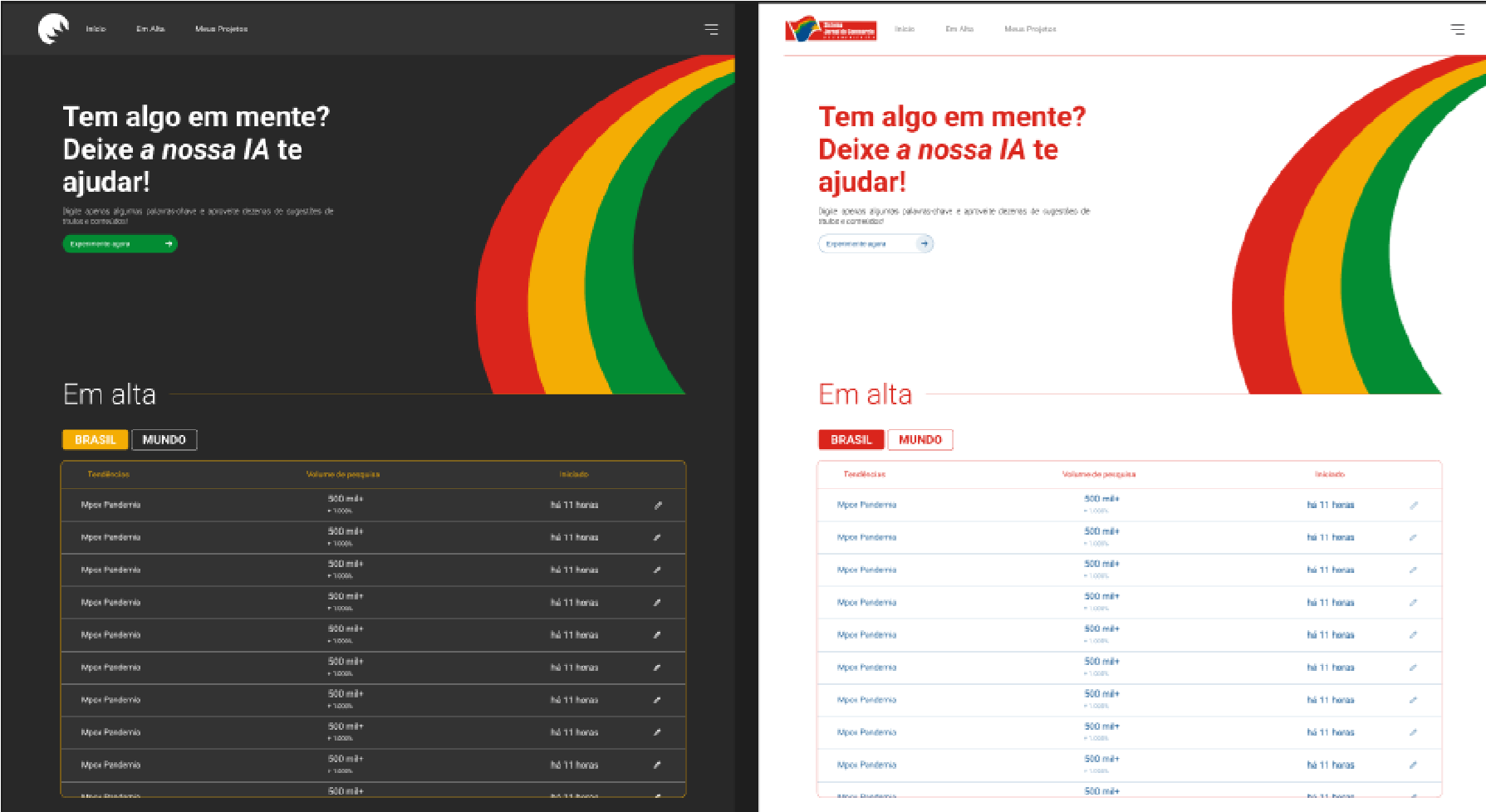}
\caption{Layout of the home page of the project (landing page), with dark and light themes.}
\label{fig:darkmode}
\end{figure}

Figure \ref{fig:darkmode} illustrates the initial interface of the system, displaying its landing page in dark and light themes, consolidating the visual identity based on the institutional colors of the SJCC and graphic elements inspired by connectivity and modernity.

%% file: text/3_Methods.tex
\subsection{System Architecture}

The architecture of the IDEIA application was designed based on principles of modularity, scalability, and loose coupling to ensure robustness in real-time data processing and facilitate integration with generative artificial intelligence models. The adopted structure uses consolidated technologies from the modern web ecosystem, with a clear separation between the presentation, business logic, and data persistence layers.

\begin{figure}[htbp]
\centering
\includegraphics[width=1\textwidth]{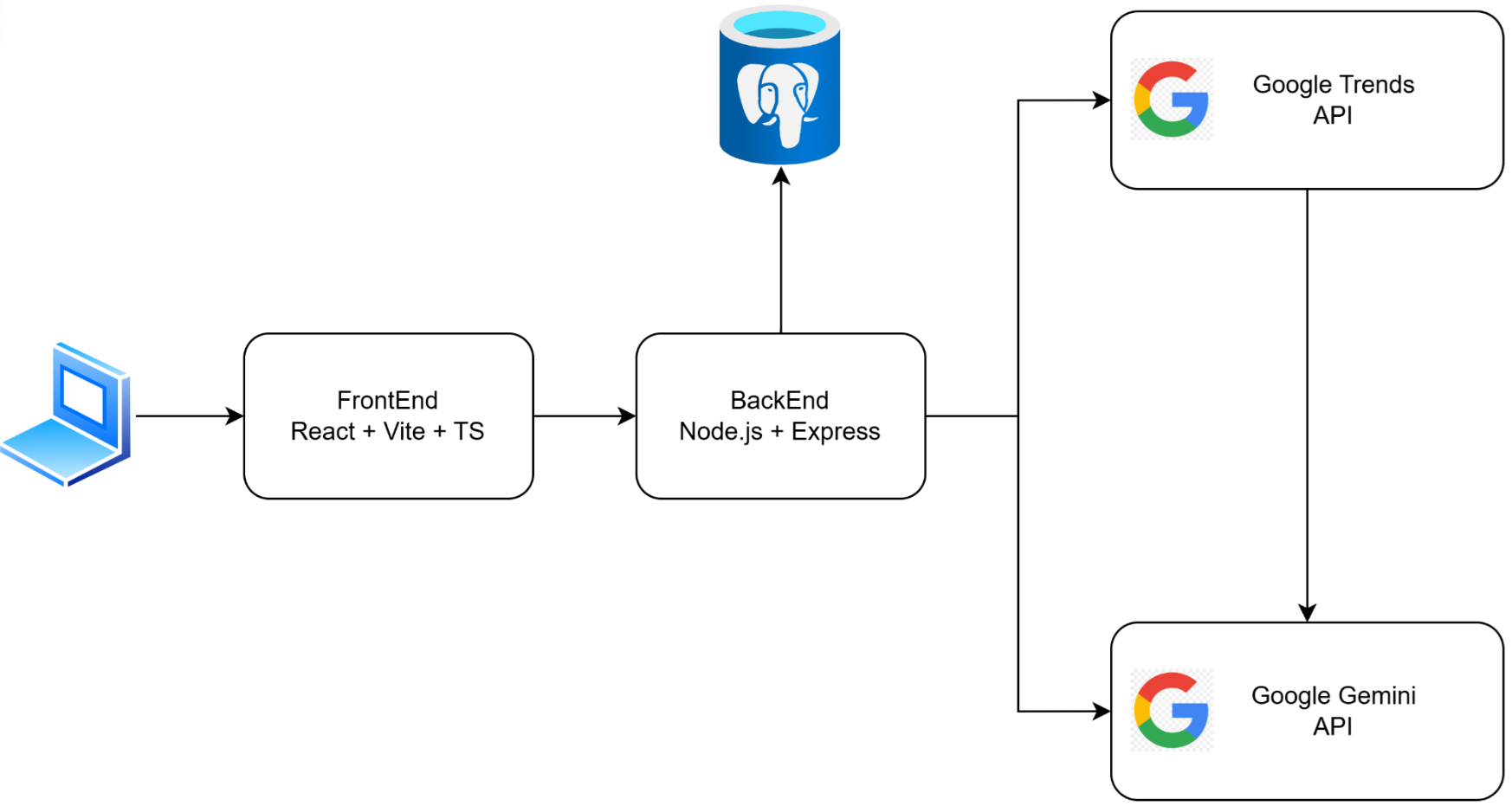}
\caption{Logical architecture of the application.}
\label{fig:arquitetura}
\end{figure}

Figure \ref{fig:arquitetura} illustrates the logical architecture of the application, highlighting the flows between the internal modules and the integrations with external APIs. The \textbf{front-end}, developed in React, Vite, and TypeScript, is responsible for the user interface and communication with the back-end. The \textbf{back-end}, in turn, is built in Node.js with Express.js, serving as an intermediary between the presentation layer, the relational database (\textbf{PostgreSQL} 16), and third-party services (\textbf{Google Trends API} and \textbf{Google Gemini API}). The infrastructure is containerized with \textbf{Docker} and hosted with support for \textbf{continuous integration} on the Vercel platform.

\subsection{Development Pipeline}

The implementation of the system followed an incremental approach, with short iterations of planning, coding, testing, and validation. The technical team was structured into specialized cells, as described below:

\begin{enumerate}
    \item \textbf{Prototyping and Interface}: Development of navigable prototypes with Figma, prioritizing responsiveness and user experience;
    \item \textbf{Front-end}: Construction of the reactive interface in React, using Vite as a bundler and Styled-Components for modular styling;
    \item \textbf{Back-end and Integrations}: Implementation of business logic and integration with external APIs via Express and official SDKs;
    \item \textbf{Persistence and Docker}: Definition of the relational model, script automation, and orchestration of Docker containers;
    \item \textbf{CI/CD and Deploy}: Configuration of the pipeline in GitHub Actions for validation, testing, and continuous publishing in Vercel.
\end{enumerate}

All steps were documented, versioned on GitHub, and accompanied by automated tests at critical points in the flow.

\subsection{Continuous Integration and Delivery (CI/CD)}

The CI/CD pipeline ensures delivery stability and agility. Table 1 summarizes the automated steps implemented.

\input{tab/tab1}

This pipeline is triggered by pushes to the main repository, allowing for the continuous release of versions with automated validations and failure traceability.

\subsection{Technical Challenges Faced}

During the development of the application, some challenges required specific technical solutions:

\begin{enumerate}
    \item \textbf{Integration with the Gemini API}: Initially, there were authentication failures and inconsistent responses. Adopting the official Google SDK solved the problem by standardizing the structure of prompts and calls.
    \item \textbf{Asynchronous State Management}: The simultaneous manipulation of multiple states in the front-end required the judicious use of hooks and attention to the life cycle of the components.
    \item \textbf{Scope Delimitation with BDD}: Using the Gherkin language allowed for alignment of expectations and enabled automated acceptance tests based on expected behaviors.
    \item \textbf{Persistence with Docker}: Initial problems in preserving data between container restarts were solved with the explicit configuration of named volumes.
\end{enumerate}

%% file: tab/tab1.tex
\begin{table}[htbp]
\centering
\caption{Trends table extracted from the Google Trends API.}
\begin{tabularx}{\textwidth}{c|c|X}
\hline
\textbf{Step} & \textbf{Tool} & \textbf{Description} \\ \hline
Continuous Integration (CI) & GitHub Actions & Runs unit tests, lint validations, and automatic builds. \\ \hline
Continuous Delivery (CD) & GitHub Actions + Vercel & Automatic publishing to Vercel after merging into the main branch. \\ \hline
Automated Tests & Jest & Unit-level tests and integration of backend components. \\ \hline
Build Monitoring & GitHub Actions & Generates build and test logs that are accessible to the entire team. \\ \hline
\end{tabularx}
\end{table}

%% file: text/4_Results.tex
The development of the IDEIA system was motivated by a specific demand from the Jornal do Commercio de Comunicação (SJCC) System: to reduce the time and effort spent on creating news stories aligned with emerging trends. Before the system was implemented, the editorial team faced a manual, time-consuming process highly dependent on the continuous reading of multiple sources. The introduction of the tool allowed this step to be automated, transferring the cognitive load of topic screening to a model based on generative artificial intelligence.

The adoption of the solution generated immediate impacts. According to reports from the editorial team, there was a significant reduction in the time needed to start writing stories, with estimated gains of close to 100\% in time savings in the story search stage. In addition to operational agility, an increase in workflow organization was observed, with greater predictability and clarity in internal processes. The system began to be seen as a support tool and a reliable interface integrated into the production routine.

\begin{figure}[htbp]
\centering
\includegraphics[width=1\textwidth]{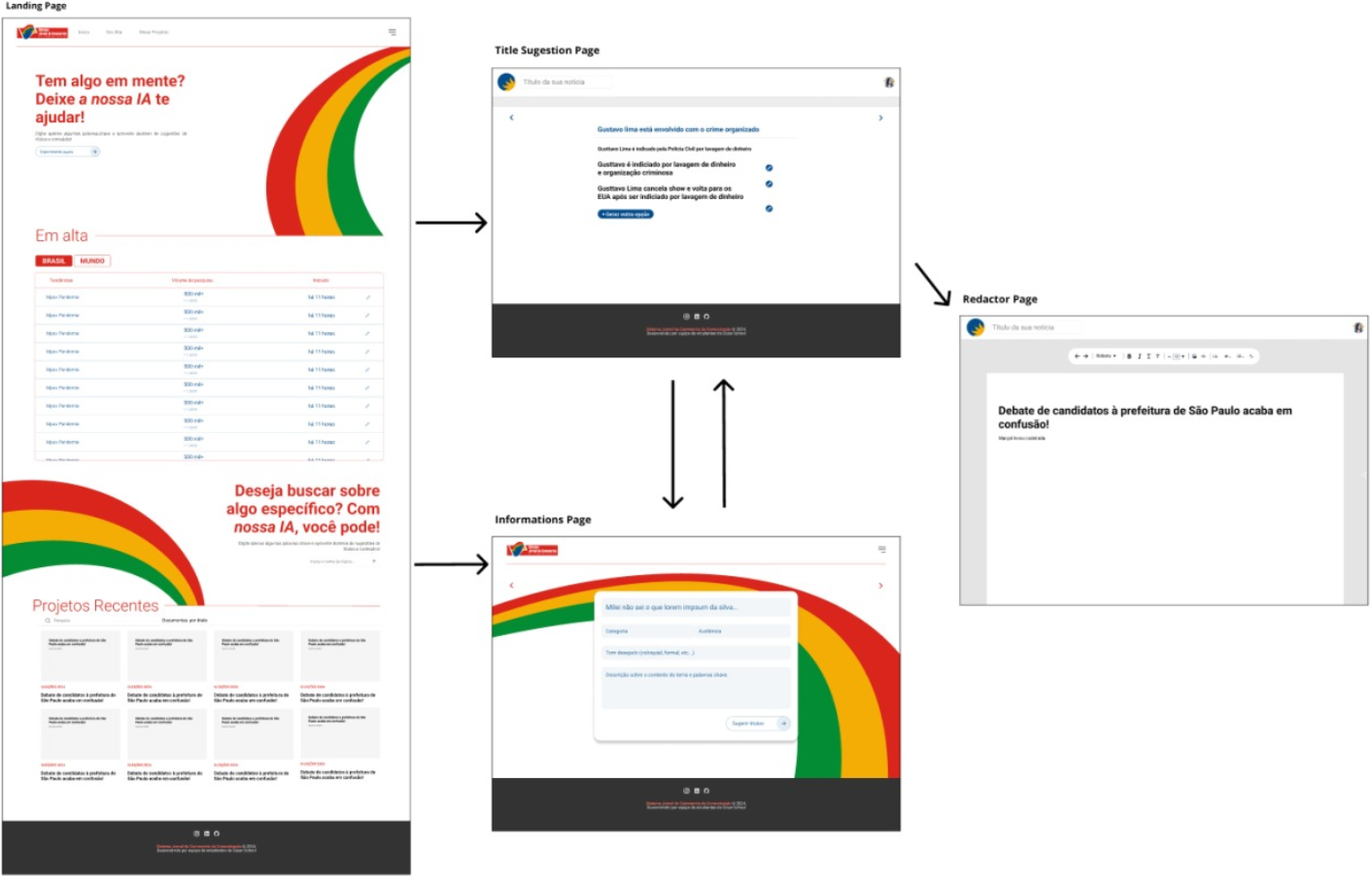}
\caption{Main screen flow of IDEIA.}
\label{fig:flow}
\end{figure}

To better illustrate the system's functioning from the user's perspective, Figure \ref{fig:flow} presents the main screen flow of IDEIA, detailing the interaction path from accessing the homepage to the suggestion, refinement, and drafting of news content. 

The process begins on the landing page, where the user selects trending topics or performs specific searches. This is followed by a step for collecting contextual information, which feeds the title suggestion engine. The flow concludes with access to the editorial environment, where the generated content can be edited and finalized. The visualization of this journey demonstrates how the system was designed to reduce friction, guide ideation, and integrate generative AI tools intuitively into the editorial routine.

\begin{figure}[htbp]
\centering
\includegraphics[width=1\textwidth]{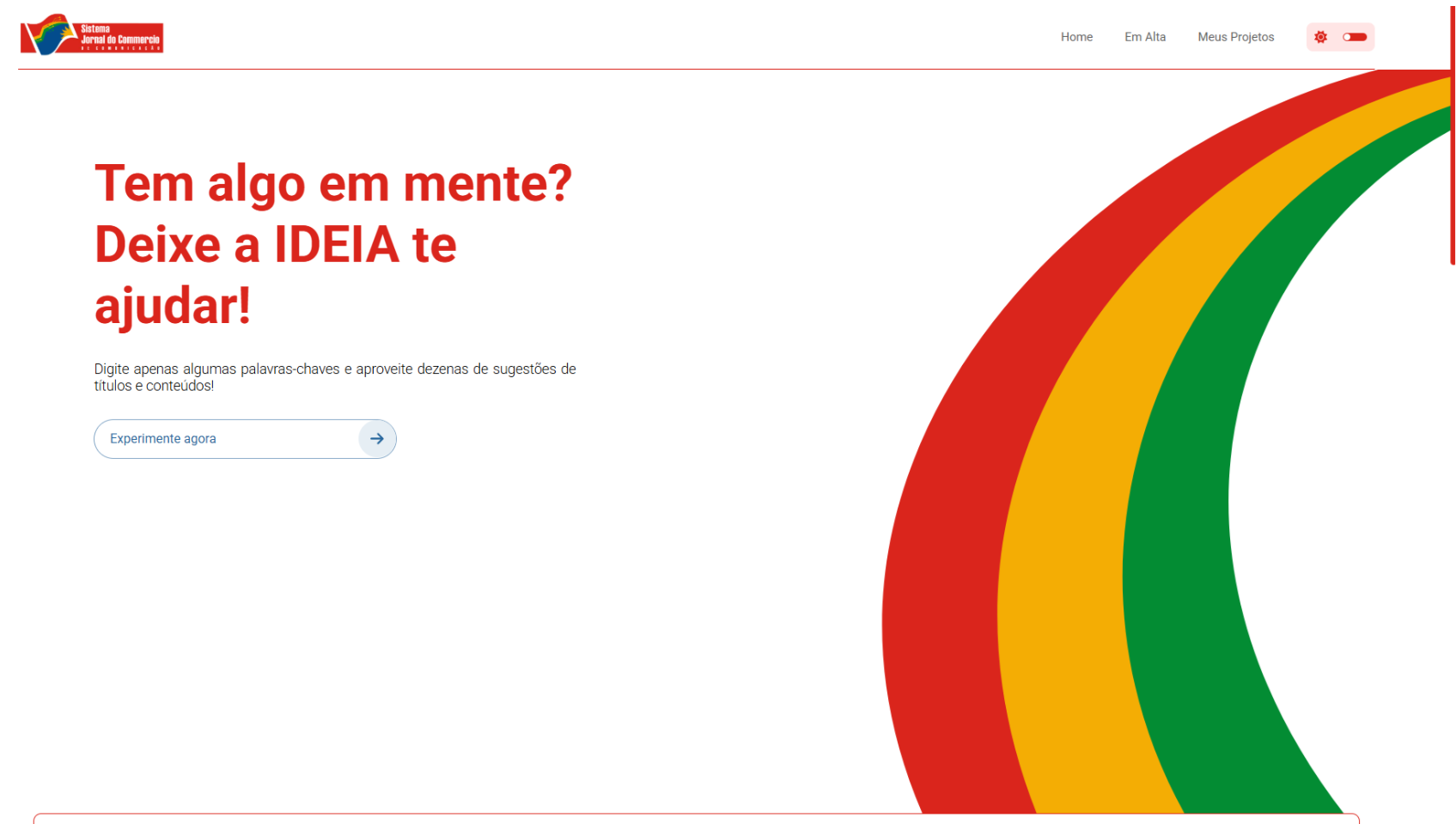}
\caption{Initial interface of the system (landing page).}
\label{fig:landingPage}
\end{figure}

Figure \ref{fig:landingPage} shows the main interface of the system, with a responsive layout, adapted to dark and light themes, respecting the institutional visual identity of SJCC and incorporating modern graphic elements that evoke connectivity, data, and innovation.

\begin{figure}[htbp]
\centering
\includegraphics[width=1\textwidth]{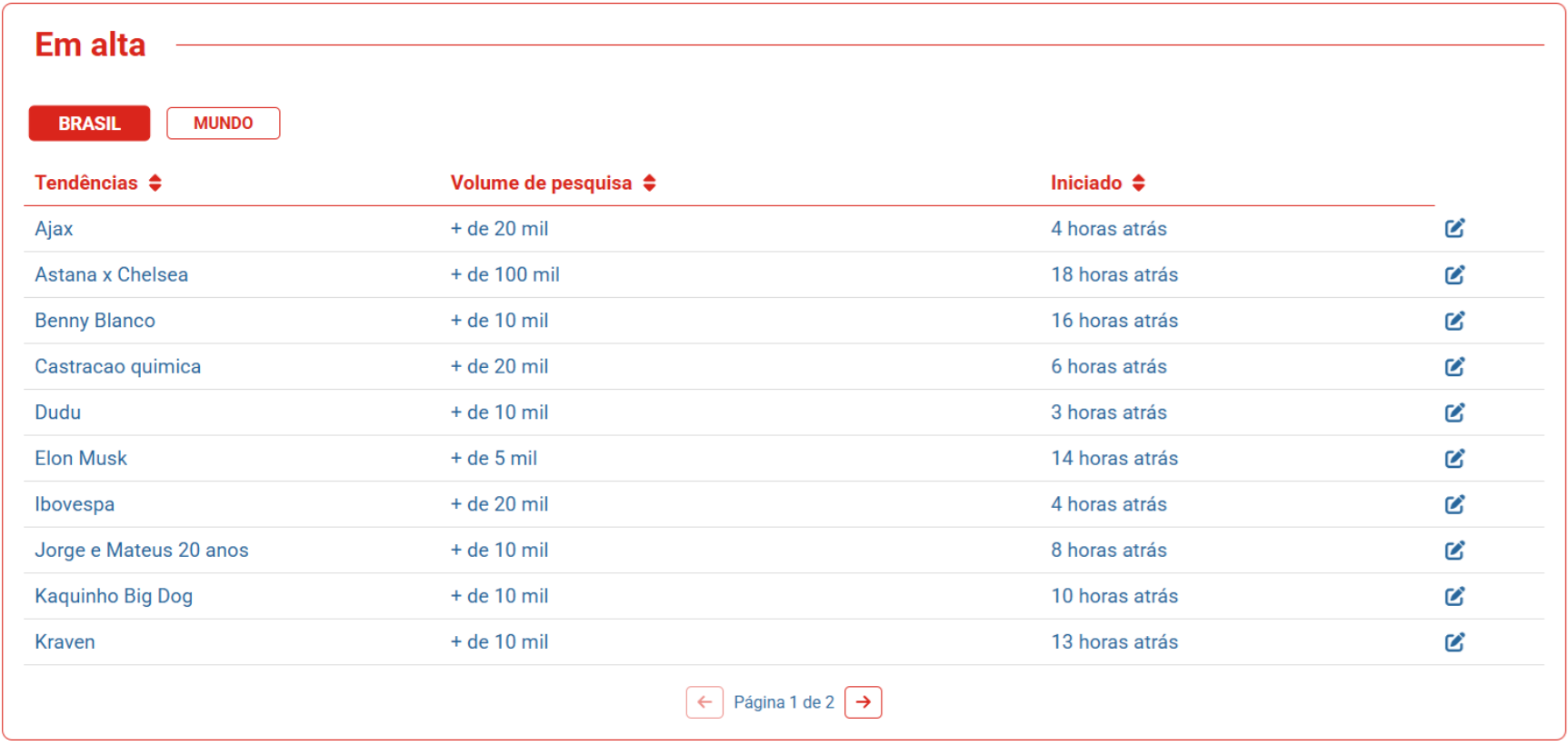}
\caption{Trends table extracted from the Google Trends API.}
\label{fig:trends}
\end{figure}

Integration with the Google Trends API was one of the features most valued by the user team. The automated extraction of data on the most searched web topics enabled the continuous visualization of emerging themes, with updates in 10-minute cycles. Figure \ref{fig:trends} illustrates the table dynamically generated by the tool, evidencing its ability to adapt to the informational context in real-time.

The solution was presented to two audiences: (i) the technical and management team of SJCC, who followed the project's progress in periodic meetings, and (ii) professors and students of the Digital Boarding Residence at Porto Digital in an event open to the academic community. In both cases, the project was received with enthusiasm and recognition. As a result, part of the team was awarded as a highlight in the Grow Up module of the OnBoard Residence due to the technical excellence and practical relevance of the product developed.

In the visual aspect, the construction of the system's identity combined traditional elements of the SJCC brand with graphic representations inspired by digital circuits. The proposal was to establish a link between innovation and tradition, conveying modernity without breaking with the institution's consolidated positioning. The project advisor, Prof. Luiz Siqueira, formally praised this strategy and highlighted the solution as "an emblematic example of innovation applied to one of society's most traditional media: journalism."

Finally, it is worth noting that, although it was initially designed to meet a specific need of the SJCC, the IDEIA system has a flexible and modular architecture, which allows it to be replicated in other departments of the company or adapted for different sectors, such as institutional communication, digital marketing, and educational content production. The scalability of the solution reinforces its potential as a reference in projects that integrate generative artificial intelligence into the information content production chain.

%% file: text/5_Conclusion.tex
The IDEIA project – Dedicated Intelligence for Writing and Advanced Inspiration – is an innovative proposal for addressing the challenges of journalistic production in dynamic digital contexts, marked by demands for agility, thematic contextualization, and informational relevance. Developed in partnership with the Jornal do Commercio de Comunicação System (SJCC), the system automated the editorial ideation stage through the integration of real-time trend data (Google Trends API) and generative language models (Google Gemini API), enabling the generation of titles and summaries that are consistent with the current informational context.

The solution demonstrated significant practical impact: it drastically reduced the time spent on thematic research, promoted greater clarity in editorial flows, and provided significant gains in productivity, with an estimated projection of up to 70\% efficiency in the agenda conception stage. These results were corroborated by qualitative feedback from SJCC teams and institutional recognition in the academic field.

From a technical standpoint, the application was built on a modern and scalable architecture, consisting of a Node.js back-end, a React front-end, orchestration via Docker, and an automated CI/CD pipeline with GitHub Actions and Vercel. This technological base supported an iterative development approach anchored in agile practices and focused on continuous value delivery.

However, important limitations are recognized. The need for prolonged empirical validation to statistically measure gains, the adaptation of the system to contexts of high editorial demand, and expansion to multilingual scenarios or with different thematic profiles are points that require future investigation. Furthermore, the integration of generative AI in creative processes raises relevant ethical issues, especially regarding algorithmic transparency, editorial responsibility, and the preservation of human authorship.

As perspectives for evolution, the following stand out: (i) the implementation of mechanisms for personalizing textual tone, (ii) the incorporation of analytical modules based on tools such as Google Analytics to measure editorial impact, and (iii) the replication of the solution in other sectors of the SJCC or journalistic organizations with a similar profile.